\documentstyle[11pt,aaspp4]{article}

\received{December 30 1997}

\slugcomment{To appear in {\it The Astrophysical Journal}}

\def\HST{{\it HST}}

\def\arcsec{\ifmmode '' \else $''$\fi}
\def\arcmin{\ifmmode ' \else $'$\fi}
\def\arcsecpoint{\ifmmode ''\!. \else $''\!.$\fi}
\def\arcminpoint{\ifmmode '\!. \else $'\!.$\fi}
\def\cc{\ifmmode {\rm cm}^{-3} \else cm$^{-3}$\fi}
\def\cl{\ifmmode {\rm cm}^{-2} \else cm$^{-2}$\fi}
\def\micron{\ifmmode \mu{\rm m} \else $\mu$m\fi}
\def\kms{\ifmmode {\rm km\,s}^{-1} \else km\,s$^{-1}$\fi}
\def\Hubble{\ifmmode {\rm km\,s}^{-1}\,{\rm Mpc}^{-1}
	\else km\,s$^{-1}$\,Mpc$^{-1}$\fi}
\def\ergsec{\ifmmode {\rm ergs\;s}^{-1} \else ergs s$^{-1}$\fi}
\def\ergscm{\ifmmode {\rm ergs\,s}^{-1}\,{\rm cm}^{-2}
	  \else ergs\,s$^{-1}$\,cm$^{-2}$\fi}
\def\ergscmA{\ifmmode {\rm ergs\,s}^{-1}\,{\rm cm}^{-2}\,{\rm \AA}^{-1}
	  \else ergs\,s$^{-1}$\,cm$^{-2}$\,\AA$^{-1}$\fi}
\def\ergscmHz{\ifmmode {\rm ergs\,s}^{-1}\,{\rm cm}^{-2}\,{\rm Hz}^{-1}
	  \else ergs\,s$^{-1}$\,cm$^{-2}$\,Hz$^{-1}$\fi}
\def\Msun{\ifmmode M_{\odot} \else $M_{\odot}$\fi}
\def\Lsun{\ifmmode L_{\odot} \else $L_{\odot}$\fi}
\def\qo{\ifmmode q_{0} \else $q_{0}$\fi}
\def\Ho{\ifmmode H_{0} \else $H_{0}$\fi}

\newcommand {\lm}{$\lambda$}

\newcommand {\lya}{Ly$\alpha$}
\newcommand {\ha}{H$\alpha$}
\newcommand {\hb}{H$\beta$}
\newcommand{\civ}{C~{\sc iv}}
\newcommand{\nv}{N~{\sc v}}

\newcommand{\oiii}{O~{\sc iii}}

\newcommand{\he}{He~{\sc ii}}
\newcommand{\ratio}{I(\lm 304)/I(\lm 1640)}

\lefthead{Zheng et al.}
\righthead{Helium Emission Line}

\begin{document}

\title{\lya/\ha\ Ratio of Singly Ionized Helium in
Quasars\footnotemark[1]}

\author{Wei Zheng}
\affil{Center for Astrophysical Sciences, The Johns Hopkins
University, Baltimore, MD 21218\\zheng@pha.jhu.edu}

\and

\author{Li-Zhi Fang}
\affil{Physics Department and Steward Observatory,
University of Arizona, Tucson,
AZ 85721\\fanglz@time.physics.arizona.edu}

\footnotetext[1]{Based on observations with the NASA/ESA Hubble Space
Telescope, obtained at the Space Telescope Science Institute, which is
operated by the Association of Universities of Research in Astronomy,
Inc., under NASA contract NAS5-26555.}

\begin{abstract}

\he\ \lya\ \lm 304/\ha\ \lm 1640 emission lines are mainly produced
by recombination, and their canonical ratio of $\sim 10$ may be
a sensitive reddening indicator. We obtain the high S/N optical
spectra of two quasars and combine them with the
far-UV spectra that show the \he\ \lm 304 emission.
For HS~1700+64, the \he\ \lm 1640 emission is not detected, and an upper
limit
to it sets the ratio greater than 20. This may not be inconsistent with
the theoretical value when all observational uncertainties are taken into
consideration. For Q0302-003, the ratio is very low, on the order of
unity.
The most plausible cause for
such a low ratio is extinction in the EUV band by very fine grains of
dust.
Q0302-003 has a prominent narrow component of $\rm FWHM \sim 2000 \ \kms$
in its major emission lines, and it appears that reddening
is associated only with the line-emitting region. We suggest that the
geometry of the line-emitting region in high-z quasars resembles that in
the low-luminosity active galaxies, with the presence of dust mostly
in the outer part.

\end{abstract}


\section{Introduction}

The first UV spectroscopic observation of a quasar (3C~273,
\cite{davidsen77}) enabled a measurement of the \lya/\ha\ ratio in an
active galactic nucleus (AGN). The low ratio of $\sim 1$ was in line with
an independent study using the composite spectrum that was derived from
various quasars (\cite{baldwin}), but was in sharp conflict with
a canonical ratio of $\sim 10$ predicted by
standard photoionization models (\cite{deo}). This ``\lya/\ha\ problem''
raised serious concerns about the validity of photoionization as the main
line-emission mechanism in AGN and
prompted extensive theoretical interest in the following years.
Improved photoionization models (\cite{ferland}, and
references therein)
invoke large column densities and moderate degrees of ionization.
In a partially ionized zone, the low escape probability for \lya\ photons
makes a high population of excited states, and
collisional excitation from these levels enhances Balmer lines.
Calculations using a reasonable AGN continuum lead to an enhancement of
Balmer lines by a factor of $\sim 2$, and hence may not fully explain the
observed low \lya/\ha\ ratio. Another explanation introduces intrinsic
reddening in the line-emitting region
(\cite{dust}). The wavelength-dependent extinction reduces the
intensities of observed UV lines, thus lowering the \lya/\ha\ ratio.
The observed Pa$\alpha$/\ha\ ratio, however, appears to be too low for
a straightforward full account as a reddening effect (\cite{rick}).
Significant evidence exists for dust in the narrow-line region of
Seyfert galaxies (\cite{deo}; \cite{n93}) and
for a decrease in extinction with increasing luminosity
toward quasars (\cite{cdz}; \cite{rudy}). However, \cite{wills} suggested
that in intermediate-redshift quasars there may be significant
reddening in the narrow-line region.

Accurate assessments of the reddening effect depend on the use of
good line pairs whose intrinsic ratios are fairly stable.
Since singly ionized helium is hydrogenic, the \he\ I(\lya\ \lm
304)/I(\ha\ \lm
1640) ratio should therefore be the same as that for hydrogen.
The \he\ emission is produced mainly by recombination,
because its excitation level of 40 eV is considerably higher than the
average thermal energy in the line-emitting region.
The wavelength of \he\ \lm 304 emission coincides with that of O$^{++}$
transition $2p^2\ ^3P_2 - 2p\ 3d\ ^3P_2^0$,
allowing Bowen fluorescence radiation (\cite{eastman}; \cite{netzer}).
This radiation mechanism, however, does not appreciably affect the \he\
ratio itself.
The \he\ \lm 304 emission should be extremely sensitive to reddening.
While there are no hurdles in the theoretical aspects, it has taken some
20 years to advance from measuring this ratio in hydrogen to that in
helium.

\section{Observations}

As a result of searches for the \he\ Gunn-Peterson effect in the
intergalactic medium, the UV spectra of several hundred quasars have been
studied with \HST. The UV spectra of most high-z quasars are cut off by
intervening Lyman-limit
systems, and only four quasars have so far been found to
exhibit a detectable flux
below the wavelength of the redshifted \he\ \lm 304 emission:
Q0302-003 (z=3.286; \cite{jakobsen}),
PKS~1935-692 (z=3.185; \cite{tytler}), HS~1700+64 (z=2.743; \cite{dkz}),
and HE~2347-4342 (z=2.885; \cite{reim2}).
The \he\ \lm 304 emission, even with a profile that is partially absorbed,
makes it possible to carry out the ``\lya/\ha'' test for ionized helium.
We used the UV data of HS~1700+64 obtained with the Hopkins Ultraviolet
Telescope (HUT) during the Astro-2 mission and that of
Q0302-003 obtained with \HST\ Goddard High-Resolution Spectrograph (GHRS)
at an improved S/N ratio and resolution (\cite{hogan}). The GHRS data were
retrieved from the \HST\ archive and
co-added.

We obtained the optical spectra of these two quasars on several occasions.
The results presented here are mainly from two observing runs in 1996 July
and November. At the 2.3m Bok telescope of Steward
Observatory, University of Arizona, we used the B\&C spectrograph with a
Loral $1200 \times 800$ CCD detector and a long slit of
$2\arcsecpoint 5$. On July 16 (UT), the observations of HS~1700+64 were
made with a grating of 600 line mm$^{-1}$ for 10,060~s,
yielding a resolution of $\sim 3.7$~\AA\ between $\sim 5400$ and 7600~\AA.
On November 4 and 6 we observed Q0302-003. On the first night,
a total exposure of 21,600 s was made using
a grating of 832 line mm$^{-1}$ that allows a spectral coverage from
$\sim 5900$ to 7600~\AA\ with a resolution of 2.4~\AA. On the other night,
the total exposure was 25,200 s, and
a grating of 400 line mm$^{-1}$ was used,
covering $\sim 5600$ and 8800~\AA\ with a resolution of $\sim 5.8$~\AA.
With
the seeing of $\sim 1\arcsecpoint 5$, all the observations were made with
a series
of exposures of 30 minutes each. On each night, low-resolution spectra
were
also obtained to cover the wavelength band near the \lya\ emission.

The data were processed and analyzed using standard {\it IRAF} tasks.
The A- and B-band absorption were corrected
with the templates derived from the standard stars. The removal
was only partial because some exposures were obtained with large air mass
without corresponding spectra of nearby template stars. Spectra taken on
different dates were combined, using exposure times as the weight.
The spectra of HS~1700+64 and Q0302-003 are displayed respectively in Fig.
1
and 2.
The spectrum of HS~1700+64 covers too small a wavelength region shortward
of the \civ\ to allow an accurate determination of the continuum level. We
used another spectrum taken in 1994 June with the same telescope
that covers wavelengths as low as 5000~\AA. The fitted continuum matches
that spectrum well. We therefore feel that the continuum used in the
fitting is appropriate.

The color excess $E_{B-V}$ was calculated as 0.05 for HS~1700+64 and 0.12
for Q0302-003, according to the estimated column density of Galactic
neutral hydrogen (\cite{nh}). Such a conversion
was based on the average Galactic dust-to-gas ratio derived by \cite{ebv},
which yields a larger correction than using the formulation of
\cite{bh}. The blue part of the \he\ \lm 304 profile is absorbed by the
intergalactic \he\ absorption, so we only fitted the red part. The fitting
results are presented in Table~1. The \lya\ measurements were made using
the spectra of lower resolution taken in the same observing runs.

For HS~1700+64, the fitted \he\ \lm 304 profile has a FWHM of
$\sim$ 12 000 \kms,
which matches  that of the \lya\ emission (FWHM $\sim$ 11 500 \kms;
\cite{reim1}).
In the optical spectrum
there is no noticeable emission feature at the expected \he\ \lm 1640
wavelength. To obtain an upper limit, we assumed a Gaussian profile
with a fixed centroid wavelength at 6138~\AA\ and a FWHM of 12 000 \kms.
The \he\ \lm 1640 flux is $(0.3 \pm 2.8) \times 10^{-15}$ \ergscmA,
yielding an \ratio\  ratio of greater than 20. Such a value
may not be very reliable as a slight shift in the continuum level may
change the ratio significantly. The current broad-band data of standard
stars may not allow a confirmation of broad features at 1\% level.
Therefore
the result may not be inconsistent with the canonical value of 10.
The emission centered around 6385 \AA\ does not correspond to known
emission features in quasars, and we tentatively linked it to
Si {\sc i} \lm 1701 (\cite{verner}), or a Fe {\sc ii} feature
(\cite{francis}).

For Q0302-003, there is a significant narrow component for the \lya\ and
\civ\
emission. Each of the \lya, \civ\ and \nv\ was fitted with a pair of
Gaussian components. The narrow component has a FWHM of $\sim 2000 $~\kms\
 and accounts for about a half of the line intensity.
The derived \ratio\ ratio is  $\sim 1.1$, significantly lower than
the predicted value with pure recombination.

\section{Discussions}

The \ratio\ ratio is quite different in these two quasars.
Indeed the line profiles in Q0302-003 are narrower, making
it easier to identify the weak \he\ \lm 1640 feature. The S/N level is
high
enough that a \he\ \lm 1640 feature should be detected even with a line
width
of $\sim$ 12 000 \kms. In Fig. 1 the profile of an assumed \he\ \lm 1640
feature is plotted, with an intensity 20\% of the \he\ \lm 304, which
should
have been detected. It appears that difference is not simply attributable
to
line widths.

The intensity of \he\ \lm 304 emission is affected by the Lyman line and
continuum absorption by numerous intervening absorbers along the line of
sight. This can be corrected if a high-resolution spectrum at longer
wavelengths yields a list of absorption lines. Our estimate, based
on the statistical result of \cite{valley}, suggests an optical depth of
0.2
at a rest-frame wavelength of 300 \AA\ for a z=3.3 quasar.
Therefore, this Lyman-Valley correction is not very significant.

The UV and optical observations are not simultaneous, and both quasars are
probably variable. A comparison of the UV spectra of HS 1700+64 obtained
between 1991 and 1995 finds a significant discrepancy in flux level, and
that
between the optical spectra taken between 1994 and 1996 shows that the
\lya\ equivalent width varies by a factor of 2. The \civ\ equivalent width
of
Q0302-003 has increased by $\sim 50\%$ as compared with the data of
Sargent, Steidel \& Boksenberg (1989). Furthermore, the photometric
quality of
our optical spectra is questionable.
A typical light loss with a small slit during an optical spectroscopic
observation is $\sim 15\%$. These factors add uncertainties to the \ratio\
ratio. The derived line ratio is also subject to the reddening
formulation.
If we use the formula of Burstein \& Heiles (1978), this ratio would be
even
lower.

We have carried out photoionization calculations (\cite{cloudy}) with
various parameters. With a broad range of the density,
column density, flux and shape of
the ionizing continuum,  the \ratio\ ratio
varies within a narrow range between 9 and 11. It is therefore not
practical to attribute the observed low value to special conditions
in the line-emitting region. Note that a part of the \he\ \lm 304 emission
may
receive a contribution  from \oiii\ \lm 305 emission that is produced
by Bowen fluorescence mechanism (\cite{eastman}). If this were the case,
the actual \ratio\ ratio would be even lower.

The low \ratio\ ratio in Q0302-003 may signal internal reddening
in the line-emitting region.
Dust grains with dimensions of $\sim 3 \times 10^{-6}$ cm are believed to
produce Galactic extinction (\cite{ext}) which generally follows a
$1/\lambda$ law. If intrinsic extinction is produced by even smaller
grains, and the wavelength dependence of the extinction law applies to
wavelengths as
short as 300~\AA, then an $E_{E - V} = 2.5$, where $E$ denotes a band
around 300 \AA, is needed to explain the
discrepancy between the observed and theoretical ratio of \ratio.
This would translate into $E_{B-V} = 0.5$. If this is the case,
significant
presence of dust may be a reality in the broad-line region of some
high-redshift quasars.

The quantitative formalism should be more complicated than that.
While the extinction curves between 1 $\mu$m and 1000~\AA\ can be
approximated
with a $1/\lambda$ law, very little is known about the extinction
properties below 1000~\AA. \cite{hawkins} and \cite{martin} calculated the
EUV extinction curve for graphite-silicate dust. Their results show
{\em decreasing} extinction from 1000~\AA\ to 100~\AA.
\cite{pei} suggested that a numerical formula can be applied to
other galaxies, possibly to those of higher redshifts, without assuming
a Galactic dust-to-gas ratio. If such extinction is real, the same
extinction affects the intensity of
other UV lines as well. The intrinsic hydrogen \lya/\ha\ ratio in these
objects may actually be higher than the observed ones. Likely, the
average ratio would be $ \sim 6$, closer than the theoretical value of
$\sim 10$. This will, in turn, help understand the classical \lya/\ha\
puzzle. If significant reddening does exist in the quasar broad-line
region, the observed \lya/\ha\ ratio may be corrected upward by an
additional factor of $\sim 2$.

Even for these two quasars,
the I(\lm 1216)/I(\lm 304) ratio is very different. In
HS~1700+64, this ratio is about 3, while in Q0302-003 it is about 30.
Significant
extinction in the EUV band can explain both the abnormal line ratios.
The I(\lm 1216)/I(\lm 304) ratio in both objects is 30 or higher,
consistent
with photoionization models. Therefore, the likely cause for the low
ratio in Q0302-003 is extinction in the broad line region by very small
grains.

It may not be coincidental that narrow line widths and possibly
significant
reddening are present in the same object. Seyfert 2 galaxies often
show a higher degree of extinction (\cite{deo}), and the narrow-line
region
in Seyfert-1 galaxies generally exhibit a more significant reddening
effect
than the broad-line region (\cite{luc}).
The spectrum of Q0302-003 shows a significant narrow component with FWHM
$\sim 2000 $~\kms\ for major emission lines.
Although this line width is not considered very narrow for Seyfert
galaxies,
it is for high-z quasars. Generally, narrow lines in high-z quasars
(\cite{sargent}) are not as common as in Seyfert galaxies. For example,
\cite{wills} found no detection of
narrow line components in their radio-loud quasars of $\rm 0.26 < z <
0.77$.
They suggested a significant reddening with $E_{B-V} \simeq 0.5$.
We suggest that the
geometry of the line-emitting region in high-z quasars resembles that in
the low-luminosity active galaxies, with the presence of dust mostly
in the outer part.

Making an analogy of Seyfert galaxies and some low-redshift quasars,
we suggest that Q0302-003 has a narrow-line region which contains
a significant amount of dust.
We suggest that the
geometry of the line-emitting region in high-z quasars resembles that in
the low-luminosity active galaxies, with the presence of dust mostly
in the outer part (\cite{luc}).

Does reddening apply to the EUV continuum?
Recent studies (\cite{n95}; \cite{bechtold}) found that the \lya/\hb\
ratio
ranges between about 1 and 40 and is approximately proportional
to $f(1216)/f(4861)$, the ratio of continuum flux
at adjacent points (\cite{bechtold}). Such a correlation may suggest a
possible reddening effect that applies to both the continuum and lines.
In such cases, the equivalent widths of concerned lines should be
fairly constant. Given the significant difference in
the equivalent widths of \he\ lines in our quasar samples, we see no
compelling reason that the continuum emission from the central source is
heavily reddened.
Significant reddening would also affect the intensities of infrared
lines, and future studies of these lines may provide additional evidence
for fine dust in the quasar environment.

\acknowledgments

Support for this work has been provided in part by NASA contract
NAS-5-27000 to the Johns Hopkins University and NASA grant
AR-5284.01-93A from the
Space Telescope Science Institute, which is operated by the Association of
Universities of Research in Astronomy, Inc., under NASA contract
NAS5-26555.
We are very grateful to P. S. Smith for his assistance during the
observations.

\clearpage

\begin{deluxetable}{cc|cc|cc}
\tablecaption{Emission Line Intensities\tablenotemark{a}}
\tablewidth{0 pt}
\tablehead{
\colhead{Line} & \colhead{Wavelength} &
\multicolumn{2}{c}{HS 1700+64} & \multicolumn{2}{c}{Q0302$-$003}
\\                 &\colhead{\AA}&
\colhead{Flux} & \colhead{EW} & \colhead{Flux} & \colhead{EW}
}
\startdata
\he\tablenotemark{b} & 304 & $63 \pm 13$& 32 &$4.9 \pm 1.0$& 8 \nl
\lya & 1216 &  $202 \pm 8 $ & 142 & $155 \pm 9$\tablenotemark{c}& 305\nl
\civ & 1549& $ 117\pm 15$ & 116 &$23.5 \pm 1.8$\tablenotemark{c} & 91 \nl
\he & 1640 & $ 0.3\pm 2.8$& 0.3 & $4.5 \pm 0.4$ & 16
\enddata
\tablenotetext{a}{Fluxes are in units of $10^{-15}
$~\ergscmA, corrected for Galactic absorption.}
\tablenotetext{b}{Total intensity, based on fit to the red part of the
profile.}
\tablenotetext{c}{A narrow component accounts for about a half of the
intensity.}
\end{deluxetable}

\clearpage

\figcaption{Optical spectrum of HS~1700+64 (z=2.743). The atmospheric
B-band absorption, marked with an Earth symbol, has been partially
corrected.
The fitted power-law continuum is plotted in dashed curve, and the
fitting windows are marked.
An assumed \he\ \lm 1640 profile, with 1/5 of the
intensity of \he\ \lm 304 line is plotted in dotted curve at the bottom,
to allow an assessment of the \he\ \lm 1640 intensity.
A tentative line identification is marked with a question mark.
\label{fig1}}

\figcaption{Optical spectrum of Q0302-003 (z=3.286). The atmospheric
A- and B-band absorption features, marked with Earth symbols, have been
partially corrected.
The fitted power-law continuum is plotted in dashed curve, and the
fitting windows are marked.
\label{fig2}}
\clearpage

\setcounter{figure}{0}
\begin{figure}
\plotfiddle{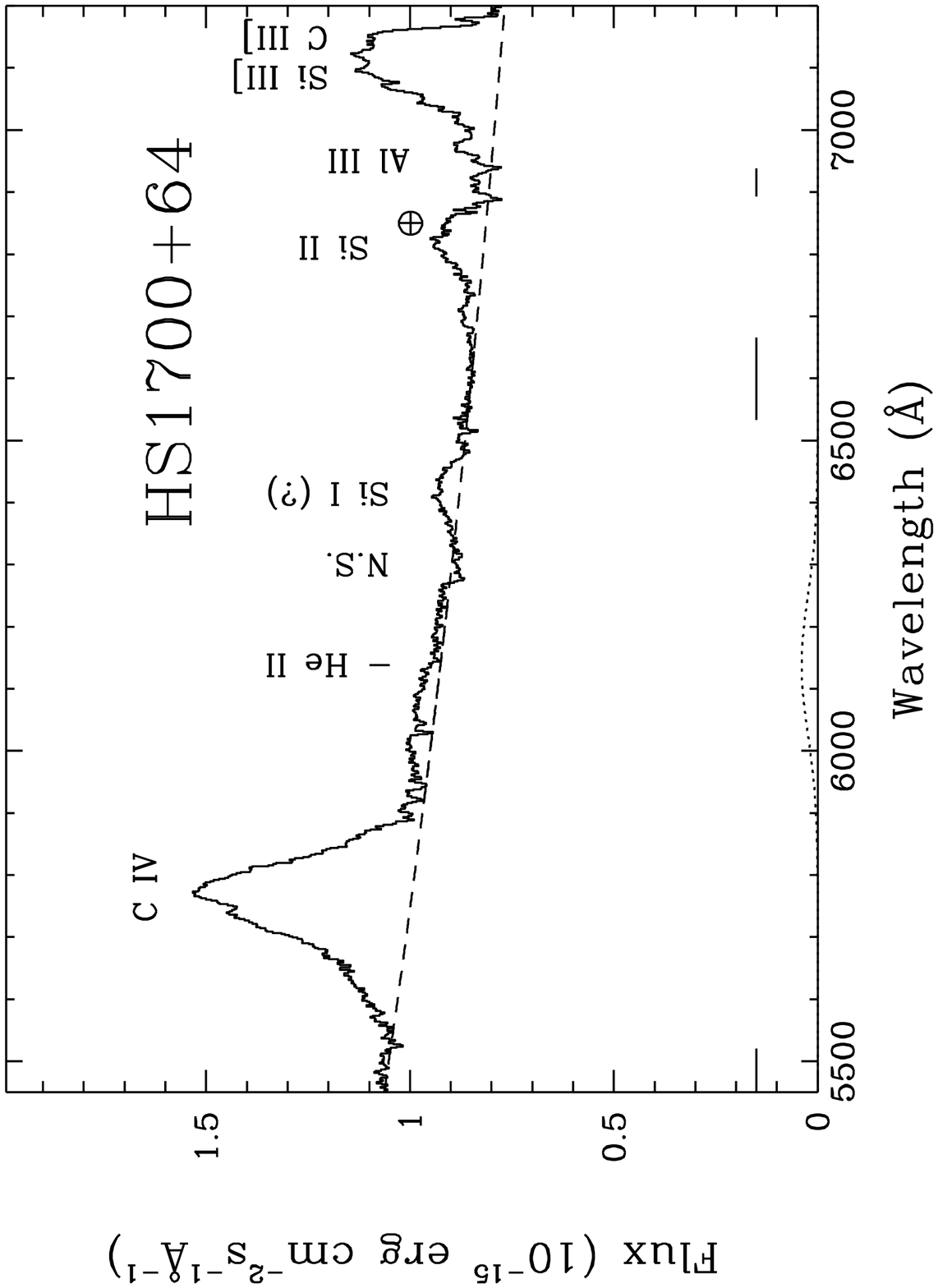}{6 in}{-90}{60}{60}{-240}{370}
\caption{~}
\end{figure}

\begin{figure}
\plotfiddle{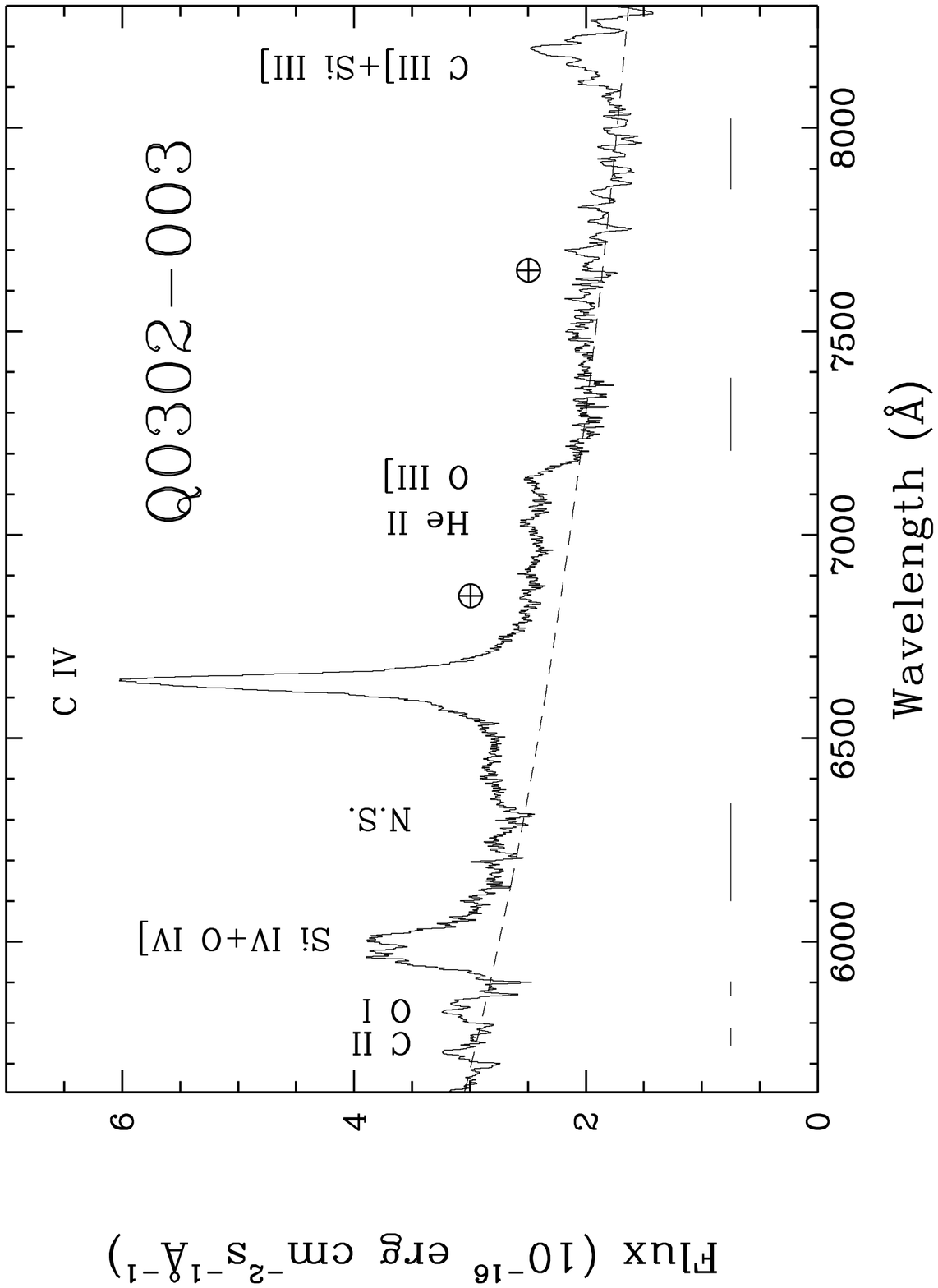} {6 in}{-90}{60}{60}{-240}{370}
\caption{~}
\end{figure}

\end{document}